\documentclass[letterpaper,english,aps]{revtex4-1}
\usepackage[LGR,T1]{fontenc}
\usepackage[latin9]{inputenc}
\usepackage{color}
\usepackage{amsmath}
\usepackage{amssymb}

\makeatletter

\DeclareRobustCommand{\greektext}{%
  \fontencoding{LGR}\selectfont\def\encodingdefault{LGR}}
\DeclareRobustCommand{\textgreek}[1]{\leavevmode{\greektext #1}}
\ProvideTextCommand{\~}{LGR}[1]{\char126#1}

\makeatother

\usepackage{babel}
\begin{document}
\title{Reply on ``Comments on Superstatistical properties of the one-dimensional
Dirac oscillator by Abdelmalek Boumali et al.''}
\author{Abdelmalek Boumali}
\email{boumali.abdelmalek@gmail.com; abdelmalek.boumali@univ-tebessa.dz}

\affiliation{Laboratoire de Physique Appliqu�e et Th�orique, ~\\
Universit� Larbi-T�bessi-T�bessa, Algeria.}
\author{Samia Dilmi}
\email{dilmisamia@gmail.com}

\affiliation{Facult� des Sciences exactes, Universit� Hama Lakhdar d\textquoteright El-Oued,~\\
B.P. 789 El-Oued 39000, Algeria}
\author{Fadila Serdouk}
\email{iserdouk@yahoo.fr;iserdouk@univ-tebessa.dz}

\affiliation{Laboratoire de Physique Appliqu�e et Th�orique, ~\\
Universit� Larbi-T�bessi-T�bessa, Algeria.}

\maketitle
In the work \textquotedblleft Comments on Superstatistical properties
of the one-dimensional Dirac oscillator by Abdelmalek Boumali et al.\textquotedblright ,
Castano-Yepes et al \citep{1} have mentioned three remarks on our
paper, and they claimed that our calculations do not have a rigorous
treatment \citep{2}. Their remarks are the following
\begin{itemize}
\item The expansion of the partition function in terms of powers of the
parameter q is incomplete.
\item Our canonical partition function is incomplete because we have ignored
all the poles located at the negative real axis.
\item Our extension to the case of Graphene is also mistaken, given the
fact that it was performed under a wrong formalism and with an incomplete
partition function.
\end{itemize}
In what follows, we will discuss their comments point by point:
\begin{itemize}
\item For the first point, the authors have not well understand our approach:
\end{itemize}
In statistical mechanics, the habitual Boltzmann factor $e^{-\beta E}$
is an essential tool used to determine thermodynamic quantities such
as the partition function $Z(\beta)$, free energy $F(\beta)$, total
energy $U(\beta)$, entropy $S(\beta)$ and specific heat $C(\beta)$,
for a given system. These quantities are defined as
\begin{equation}
F=-\frac{\text{ln}Z}{\beta},\,U=-\frac{\partial\text{ln}Z}{\partial\beta},\label{eq:1}
\end{equation}
\begin{equation}
\frac{S}{k_{B}}=\text{ln}Z-\beta\frac{\partial\text{ln}Z}{\partial\beta},\,\frac{C}{k_{B}}=\beta^{2}\frac{\partial^{2}\text{ln}Z}{\partial\beta^{2}}.\label{eq:2}
\end{equation}
As we know, superstatistics is a branch of statistical mechanics or
statistical physics devoted to the study of non-linear and non-equilibrium
systems. Generally, the applicability of these equations in the non-equilibrium
system is not valid. To extend all well known formulae of normal statistical
mechanics to the case of superstatistics, are restricted by the following
conditions: First, superstatistics is characterized by using the superposition
of multiple different statistical models to achieve the desired non-linearity.
More precisely, superstatistics assumes that the correct ensemble
is not canonical, but a superposition of canonical ensembles at different
(inverse) temperatures weighted by a factor $f(\beta)$. Thus, the
superstatistics denoted by symbol $B(E)$, allows the infinite types
of system\textquoteright s distribution with respect to $E$, once
the fluctuating distribution $f\left(\beta\right)$ is given. The
generalized Boltzmann factor $B(E)$ is based on three crucial premises;
(i) a system is partitioned into cells that can be considered to be
reached an equilibrium locally, which is characterized by a single
\textgreek{b}, (ii) its statistical factor is Gibbsian and (iii) the
separation between two time scales is adequate, that is, the time
for approaching to each local equilibrium state is much faster than
that for varying $f(\beta)$. This last criterion means that the framework
of the theory of superstatistics regards the existence of temporally
local equilibrium within each of the cells that subdivide a non-equilibrium
thermodynamic system.

Also, in order to understand well our argument, we summarize the main
results obtained by Beck \cite{3} and Tsallis et al\cite{4} about
the applicability of these equations (Eqs. (\ref{eq:1}) and (\ref{eq:2})):
Beck \cite{3} in his paper untitled ``Generalized statistical mechanics
for superstatistical systems'' developed a generalized statistical
mechanics formalism for superstatistical systems, by mapping the superstatistical
complex system onto a system of ordinary statistical mechanics with
modified energy levels. He shows, following the restriction described
above, that it is possible to do ordinary statistical mechanics for
this superstatistical non-equilibrium system, with all the known formulae.
An interesting remark can be make here: Beck introduce introduced
a universal parameter $q$ for any superstatistics, not only for Tsallis
statistics: this parameter is given by the following relation
\begin{equation}
q=\frac{\left\langle \beta^{2}\right\rangle }{\left\langle \beta\right\rangle ^{2}}.\label{eq:3}
\end{equation}
If there are no fluctuations of $\beta$ at all, we obtain $q=1$
as required.

Tsallis et al\cite{4} in their paper on the \textquotedblleft{}
Constructing a statistical mechanics for Beck-Cohen superstatistics\textquotedblright ,
strengthen the idea that the statistical-mechanical methods can be
in principle used out of equilibrium: More specifically, they used
them in equilibrium in two cases (i) in the $t\rightarrow\infty$
limit of non-interacting or short-range interacting Hamiltonians,
as well as (ii) in the $limN\rightarrow\infty$ $lim\,t\rightarrow\infty$
of long-range interacting many-body Hamiltonian systems: this explication
reinforced well our restriction describe above about using the law
of ordinary statistical mechanics in the case of the superstatistics.
Thus, following all these arguments, instead a q-logarithm, we choose
the normal logarithm. 

A remark about introducing the q-algorithm here can be made: when
we use a q-logarithm, the algebra of our problem is deformed, and
it follows the theory of q-calculus with the following q-sum and q-product
definitions \cite{5,6}
\begin{equation}
x\otimes_{q}y=\left(x^{1-q}+y^{1-q}-1\right)_{+}^{\frac{1}{1-q}},\,\left(x>,\,y>0\right)\label{eq:4}
\end{equation}
\begin{equation}
x\oplus_{q}y=x+y+\left(1-q\right)xy.\label{eq:5}
\end{equation}

\begin{itemize}
\item Now, for the second point, our canonical partition function is incomplete
because we have ignored all the poles located at the negative real
axis.
\end{itemize}
We start with our partition function \cite{2}
\begin{equation}
Z=\left(1+\frac{a}{2}\left\langle \beta\right\rangle ^{2}\frac{d^{2}}{d\left\langle \beta\right\rangle ^{2}}+\frac{a^{2}}{3}\left\langle \beta\right\rangle ^{3}\frac{d^{3}}{d\left\langle \beta\right\rangle ^{3}}\right)\sum_{n}e^{-\sqrt{2r}\left\langle \beta\right\rangle mc^{2}\sqrt{n+\frac{1}{2r}}}\label{eq:6}
\end{equation}
By using the Mellin transformation, the sum in Eq. (\ref{eq:6}) is
transformed into integral as follows:
\begin{align}
\sum_{n}e^{-\sqrt{2r}\left\langle \beta\right\rangle mc^{2}\sqrt{n+\frac{1}{2r}}} & =\frac{1}{2\pi i}\int_{C}ds\left(\left\langle \beta\right\rangle mc^{2}\sqrt{2r}\right)^{-s}\sum_{n}\left(n+\frac{1}{2r}\right)^{-\frac{s}{2}}\Gamma\left(s\right)\nonumber \\
 & =\frac{1}{2\pi i}\int_{C}ds\left(\left\langle \beta\right\rangle mc^{2}\sqrt{2r}\right)^{-s}\zeta_{H}\left(\frac{s}{2},\frac{1}{2r}\right)\Gamma\left(s\right),\label{eq:7}
\end{align}
$\Gamma\left(s\right)$ and $\zeta_{H}\left(\frac{s}{2},\alpha\right)$
are respectively the Euler and Hurwitz zeta function \cite{7}. As
we know, the Gamma Function has simple poles at the non-positive integers,
and their residues are 
\begin{equation}
\text{Re}\left\{ \Gamma,-n\right\} =\frac{\left(-1\right)^{n}}{n!}.\label{eq:8}
\end{equation}
On the other hand, the Hurwitz zeta function $\zeta_{H}\left(s,\alpha\right)=\sum_{n=0}^{\infty}\frac{1}{\left(n+\alpha\right)^{s}}$.
It is a series that converges only when $\mathcal{R}\left(s\right)>1$
and $\mathcal{R}\left(\alpha\right)>0$. It can be extended by analytic
continuation to a meromorphic function defined for all complex numbers
s with $s\neq1$. At $s=1$ it has a simple pole with residue 1 \cite{7}.
Applying the residues theorem on Eq. (\ref{eq:7}) gives
\begin{equation}
\sum_{n}e^{-\sqrt{2r}\left\langle \beta\right\rangle mc^{2}\sqrt{n+\frac{1}{2r}}}=\frac{1}{2r\left(\left\langle \beta\right\rangle mc^{2}\right)^{2}}+\sum_{n}\zeta_{H}\left(-\frac{n}{2},\alpha\right)\frac{\left(-1\right)^{n}}{n!}.\label{eq:9}
\end{equation}
Here, as mentioned by the authors \cite{8,9}, we have only two poles
at $s=0$ (pole of $\Gamma$) and $s=2$ (pole of $\zeta_{H}\left(\frac{s}{2},\alpha\right)$.
For the other poles of the function $\Gamma$, $\zeta_{H}$ does not
converges. For a special case of $s=0$, we have applied the following
properties \cite{10}
\begin{equation}
\zeta_{H}\left(-n,\alpha\right)=-\frac{B_{n+1}}{n+1}\label{eq:10}
\end{equation}
where $B_{n}$ are the Bernoulli polynomials. Here The first few Bernoulli
polynomials: 
\begin{equation}
B_{0}\left(x\right)=1,\label{eq:11}
\end{equation}
\begin{equation}
B_{1}\left(x\right)=x-\frac{1}{2},\label{eq:12}
\end{equation}
\begin{equation}
B_{2}\left(x\right)=x^{2}-x+\frac{1}{2},\label{eq:13}
\end{equation}
\begin{equation}
B_{3}\left(x\right)=x^{3}-\frac{3}{2}x^{2}+\frac{1}{2}x.\label{eq:14}
\end{equation}
\begin{equation}
\vdots\label{eq:15}
\end{equation}
In particular, the relation holds for $n=0$ and one has
\[
\zeta_{H}\left(0,\alpha\right)=-B_{1}=\frac{1}{2}-\alpha.
\]

\begin{itemize}
\item Finally, according to our arguments described above, their comments
on the extension of our study to the case Graphene are not correct.
\end{itemize}

\end{document}